# Stability of boron-doped graphene/copper interface:

# DFT, XPS and OSEE studies

D. W. Boukhvalov[1,2], I. S. Zhidkov[3], A. I. Kukharenko[3], A. I. Slesarev[3], A. F. Zatsepin[3],

S. O. Cholakh[3] and E. Z. Kurmaev[3,4]

[1]*Department of Chemistry, Hanyang University, 17 Haengdang-dong, Seongdong-gu, Seoul 04763, Korea*
[2]*Theoretical Physics and Applied Mathematics Department, Ural Federal University, Mira Street 19, 620002 Yekaterinburg, Russia*
[3]*Institute of Physics and Technology, Ural Federal University, 620002 Yekaterinburg, Russia*
[4]*M.N. Mikheev Institute of Metal Physics, Russian Academy of Sciences, Ural Branch, 620108 Yekaterinburg, Russia*

*Two different types of boron-doped graphene/copper interfaces synthesized using two different flow rates of Ar through the bubbler containing the boron source were studied. X-ray photoelectron spectra (XPS) and optically stimulated electron emission (OSEE) measurements have demonstrated that boron-doped graphene coating provides a high corrosion resistivity of Cu-substrate with the light traces of the oxidation of carbon cover. The density functional theory calculations suggest that for the case of substitutional (graphitic) boron-defect only the oxidation near boron impurity is energetically favorable and creation of the vacancies that can induce the oxidation of copper substrate is energetically unfavorable. In the case of non-graphitic boron defects oxidation of the area, a nearby impurity is metastable that not only prevent oxidation but makes boron-doped graphene. Modeling of oxygen reduction reaction demonstrates high catalytic performance of these materials.*

# 1. Introduction

Graphene-based materials were extensively studied last years due to their special for carbon allotropes 2D-structure and extraordinary mechanical, electrical and chemical properties [1-5]. To further improve their usefulness, many methods have been suggested among them, the doping with heteroatoms (such as B, N, P, and S) has been shown to be an effective way to modify the electrochemical properties and to enhance their capacitive performances [6-8]. Particularly, the doping of graphene by boron seems to be very attractive because of their close atomic sizes that should avoid the formation of structural defects such as carbon vacancies which are found, for instance, under doping of graphene by nitrogen [9]. According to DFT calculations, the energy barrier to incorporate the boron atoms to graphene lattice is lower compared with that of nitrogen [10]. Incorporation of boron atoms into an aromatic carbon framework like carbon nanotubes [11] and graphene [12] offers a wide variety of functionality in, for example, chemical sensing [13, 14], nanoelectronics [15, 16], photocatalysis [17] and battery electrodes [18]. On the other hand, the chemical stability of boron-doped graphene coatings is not studied yet although boron has been found to be unique and efficient dopant for improving the oxidation resistance of graphite [19-22]. In the present paper, we have studied oxidation resistance of copper coated by boron-doped graphene using XPS and OSEE spectroscopy. The obtained results are compared with DFT calculations of formation energies of a different configuration of structural defects in boron-doped graphene and could also be used for the understanding of chemical properties of boron-doped graphene.

# 2. Experimental and Calculation Details

The samples of B-doped graphene B-Gr50 and B-Gr100 were synthesized at Clemson University on Cu substrates using two different flow rates of Ar through the bubbler containing the boron source (50 sccm and 100 sccm respectively), keeping the total flow rate of Ar at 450 sccm through the substrate (Fig. 1).

X-ray photoelectron spectra (XPS) were measured using a PHI 5000 Versa Probe XPS spectrometer (ULVAC Physical Electronics, USA) based on a classic X-ray optic scheme with a hemispherical quartz monochromator and an energy analyzer working in the range of binding energies from 0 to 1500 eV. Electrostatic focusing and magnetic screening were used to achieve an energy resolution of $\Delta E \leq 0.5$ eV for the Al $K_\alpha$ radiation (1486.6 eV). An ion pump was used to maintain the analytical chamber at $10^{-7}$ Pa, and dual channel neutralization was used to compensate local surface charge generated during the measurements. The XPS spectra were recorded using Al $K_\alpha$ x-ray emission - spot size was 200 μm, the x-ray power delivered at the sample was less than 50 W, and typical signal-to-noise ratios were greater than 10000:3.

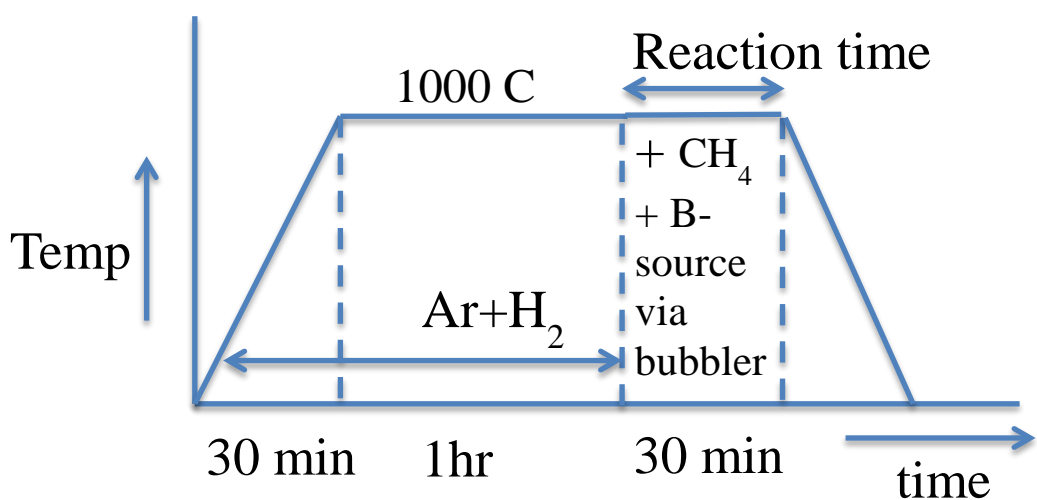

**Figure 1.** Schematic diagram of the synthesis procedure.

The optically stimulated electron emission (OSEE) measurements were carried out on experimental spectrometer ASED-1. Optical stimulation of the surface of the sample was carried out using the grating monochromator MSD-2, quartz optical system, deuterium lamp DDS-400 and secondary electron multiplier VEU-6 as electron detector. The measurements were carried out in the oil-free vacuum of $10^{-4}$ Pa at room temperature.

We used density functional theory (DFT) implemented in the pseudopotential code SIESTA, [23] as in our previous studies of similar graphene-based systems [24-25]. All calculations were performed using the local density approximation (LDA) with spin-polarization [26] which is provides better agreement with experiments [27]. During the optimization, the ion cores were described by norm-conserving non-relativistic pseudo-potentials [28] with cut off radii 1.25, 1.14, 1.59 a.u. for C, O, and B, respectively, and the wave functions were expanded with localized orbitals and a double-ζ plus polarization basis set for other species. The atomic positions were fully optimized, and optimization of the force and total energy was performed with an accuracy of 0.04 eV/Å and 1 meV, respectively. All calculations were carried out with an energy mesh cut-off of 300 Ry and a *k*-point mesh of 8×6×2 in the Monkhorst-Pack scheme [29]. A rectangular graphene supercell of 48 carbon atoms over 4 layers (24 atoms each) slab of (111) surface of fcc lattice of copper was used [24] with a single boron atom in the supercell which corresponds to an impurity concentration around 2% Used model provides separation between boron impurities more than 1 nm (four graphene lattice parameters). Thus, boron impurities can be discussed as independent and the model can be used also for description of the lower concertation of the impurities.

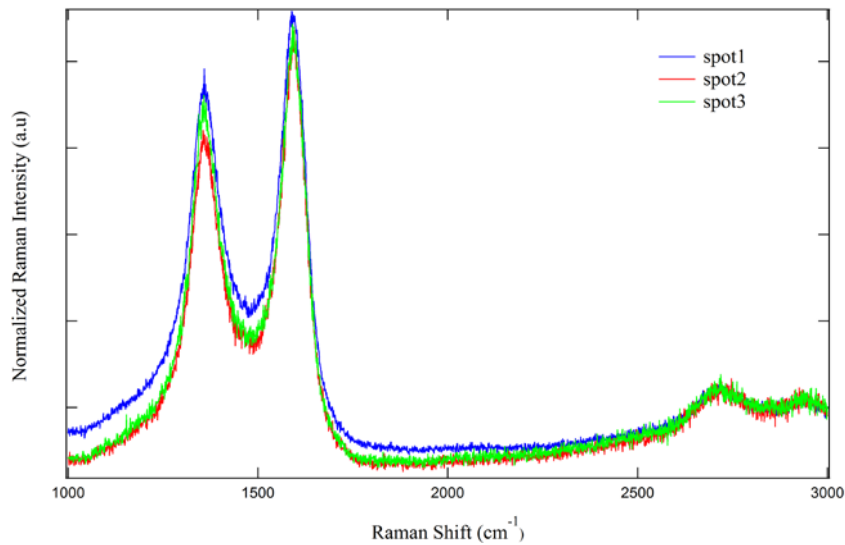

**Figure 2.** Raman spectra of boron-doped graphene.

## 3. Results and Discussions

### 3.1. Raman and XPS Measurements

The incorporation of boron into graphene lattice is confirmed by Raman spectra measurements (Fig. 2). The Raman spectra are characterized by two prominent peaks at 1340 and 1598 cm$^{-1}$, which correspond to the D and G modes, respectively. The G band originated from the doubly degenerate Brillouin zone center $E_{2g}$ phonon mode and associated with sp$^2$-hybridized carbon networks. The dispersive D band correlates with sp$^3$ hybridized carbon atoms as it requires a defect for its activation by double resonance, thus indicating the presence of lattice defects and distortions induced by boron impurity atoms [30, 31].

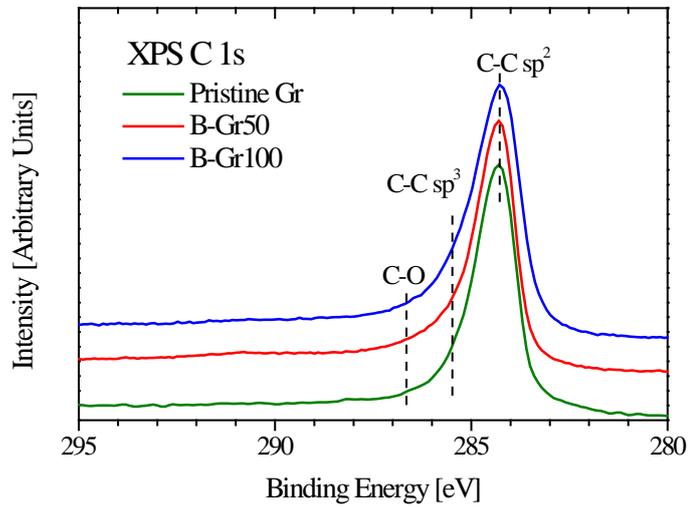

**Figure 3.** XPS C 1s of boron doped graphene/Cu interface.

This conclusion is confirmed by XPS C 1s measurements of boron doped graphene/Cu interface (Fig. 3) which show the presence both sp$^2$ and sp$^3$-contributions from graphene sheet and carbon defects, respectively, and also small oxidation of carbon atoms due to the formation of C-O functional groups [32].

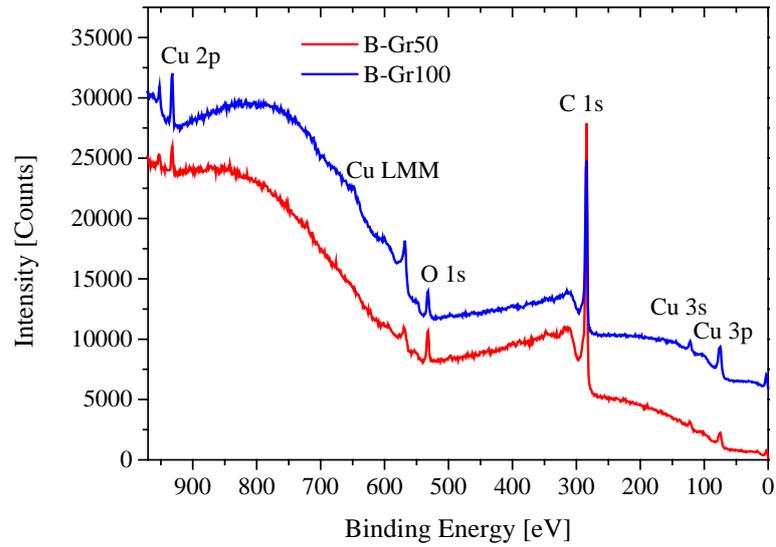

**Figure 4.** XPS survey spectra of boron doped graphene/Cu interface.

XPS survey spectra (Fig. 4) show presence only carbon, copper and oxygen lines. The boron signal is not detected. It can be due to the low doping level of boron – less than detection limit of XPS (~0.1 atom %) [33].

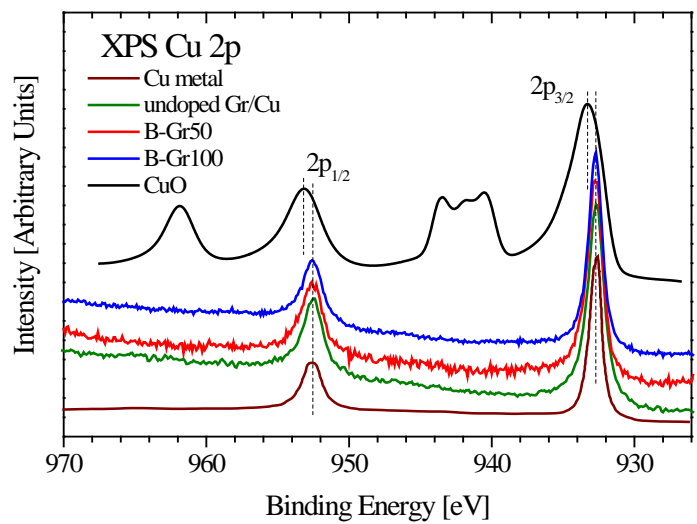

**Figure 5.** XPS Cu 2p spectra of boron doped graphene/Cu interface.

XPS Cu 2p spectra of B-doped graphene/Cu interface are found to be very similar to undoped graphene/Cu case and Cu metal and quite different concerning that of CuO (see Fig. 5). The absence of additional features in XPS specters demonstrates that neither carbon nor boron impurities do not penetrate to the copper substrate.

## 3.2. Optical properties

The results of OSEE measurements obtained by using the equation (1) [34] are reported in Fig. 6 and Table I.

$$I = A(h\nu - \varphi)^n, \qquad (1)$$

where, $I$ is OSEE intensity; $A$ - parameter approximation (scaling factor); $h\nu$ – photon energy; $\varphi$ - work function; $n$ - coefficient characterizing the type of interband (direct or indirect) electron transitions (Valence Band → Conduction Band → Vacuum Level). Parameter $n$ has the value 1; 1.5; 2; 2.5, depending on the type of optically stimulated electron transition.

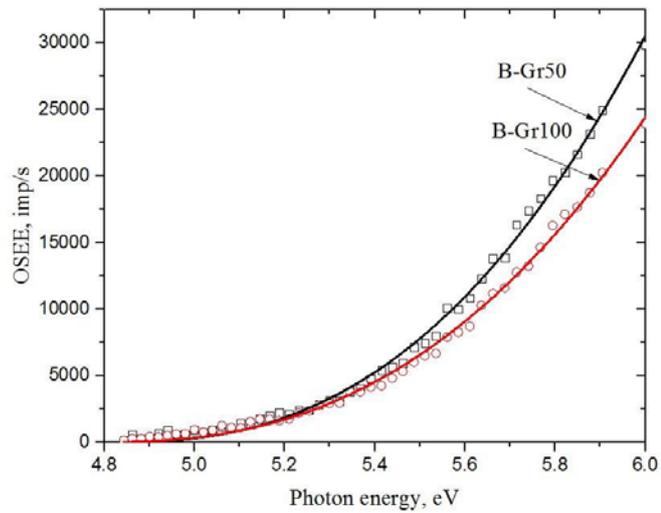

**Figure 6.** Optically stimulated electron emission from B-Gr50 and B-Gr100 (dots – experiment; curves – approximation).

**Table I.** The OSEE processing results for B-Gr50 and B-Gr100 samples.

| Sample | $\varphi$, eV | n | A | $R^2$ correlation |
|---|---|---|---|---|
| B-Gr50 | 4.80±0.02 | 2.5 | 19225 | 0.996 |
| B-Gr100 | 4.76±0.01 | 2.5 | 14028 | 0.996 |

In Tab. I we report the values of parameters $\varphi$ and $n$ which have been found by the approximation of the experimental dependences. Analysis of the data in the table leads to the following conclusions. The values of n=2.5 which are found for samples under investigation are typical for non-direct optical transitions [34] which means that process of optical excitation of electrons (Valence Band→Conduction Band→Vacuum Level) is realized with the participation of phonon sub-system. Work function φ provides the process of ionization of electronic states of sample surfaces. The $\varphi$-values reported in the Table I (4.7 -4.8 eV) are very similar for both samples and very close to that of metallic copper (4.5 eV). On the other hand, the measured $\varphi$-values are essentially different concerning those of Cu-oxides ($\varphi$ = 5.15 – 5.34 eV). [35] Therefore according to results of OSEE measurements, we also have not found any traces of copper oxidation in boron-doped graphene/Cu interfaces.

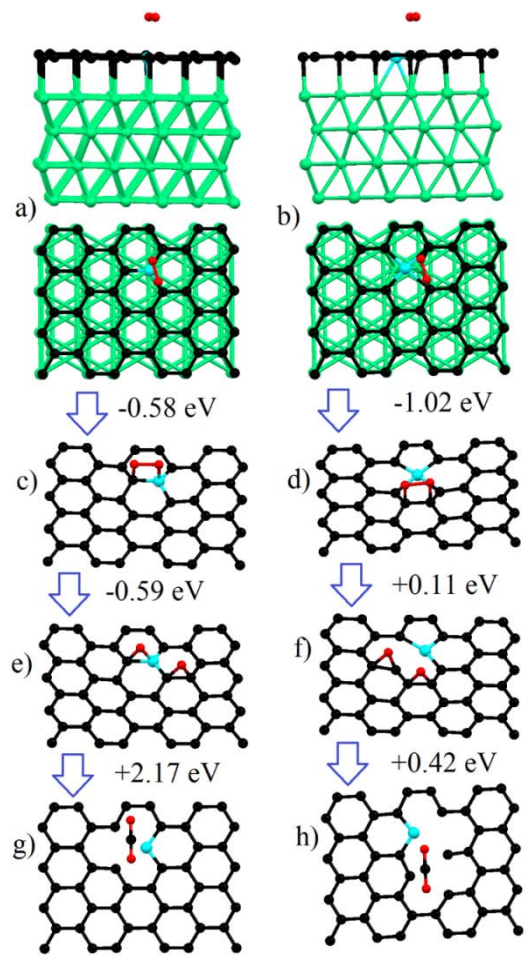

**Figure 7.** Optimized atomic structures and formation energies for step-by-step oxygen adsorption and formation of carbon vacancy in the vicinity of graphitic (left) and non-graphitic (right) boron defects. Note: on panels, c-h atoms of the copper substrate are omitted for clarity.

**Table II.** Graphene-substrate distances (in Å) and binding energies (meV/atom) for two studied types of boron defects and pure graphene on Cu.

| Configuration | Graphene-Substrate Distance | Binding Energy |
|---|---|---|
| Pure graphene on copper | 3.21 [27] | 35 [27] |
| Pure graphene on copper | 2.08 [27] | 123 [27] |
| Graphitic B-defect (Fig. 7a) | 2.43 | 121 |
| Non-graphitic B-defect (Fig. 7b) | 2.36 | 139 |

### 3.3. Modeling of the atomic structure

For the understanding of the effect of boron doping on protective properties of graphene, we performed the set of DFT modelings. At the first step, we examine atomic structure and stability of boron-doped graphene on copper substrate. Results of the calculations (Tab. II) demonstrate that graphene substrate distances and binding energies are closer to the calculated values for graphene on nickel than on copper substrate. Huge difference between nickel and copper substrates caused by the mismatch in translation vectors of graphene lattice and (111) surface of discussed metals. In the case of nickel substrate there is almost perfect coincidence of the translation vectors that provide robust interactions between graphene and metal. In the case of copper substrate small but visible mismatch of translation vectors [36,37] provides weakening of graphene-metal bonds and increasing of graphene-metal distance. The presence of boron impurity leads changes of atomic positions of carbon atoms that eliminate lattice mismatch between graphene and copper substrate. Thus, boron doping provides increasing of adhesion of graphene on copper substrate.

### 3.4. Evaluation of Chemical Stability

In recent works [38,39] the high energy barriers for the penetration of the oxygen directly through perfect graphene to the copper substrate are discussed. On the other hand, the process of oxidation of graphene provides the formation of vacancies in graphene sheets with the extraction of $CO_2$ molecule which can be permeable for oxygen. [40] We performed the modeling of the step-by-step interaction of boron-doped graphene on copper with molecular oxygen. At the first step, we checked the physisorption of an oxygen molecule (Fig. 7a,b). At the second step, we modeled the adsorption with activation (Fig. 7c,d) and decomposition to the two epoxy groups (Fig. 7e,f). At the last step, we studied the formation of carbon vacancy (Fig. 7g,h) in graphene on copper in the vicinity of two types of boron defects. The first type is the graphitic type of substitution when boron atom substitute carbon atom (Fig. 7a). In the case of boron non-graphitic defect the substitution of the pair of carbon atoms by single boron impurity takes place (Fig. 7b). In the literature, this type of defect discusses is found to be energetically favorable in the case of nano-graphenes. [41] We can also discuss this non-graphitic boron defect as the elimination of Stone-Walles and similar defects when boron impurity substitutes two displaced carbon atoms and speculate that boron impurities sit on the grain boundaries with multiple Stone-Walles and similar defects. [42]

Results of the calculations demonstrate that in contrast to pure free-standing graphene [43] and graphene on copper [41] the activation of the oxygen in the vicinity of boron defects is an exothermic process. This type of oxygen activation in graphene in the vicinity of boron impurity could be the explanation of experimentally detected catalytic properties of boron doped graphene. [42-45] In the case of graphitic defects the further decomposition of O-O pair is also energetically favorable that could provide an oxidation of graphene in the vicinity of graphitic defects. This oxidation could be a source of minor oxygen contribution to XPS spectra (Fig. 3), but the energy cost of formation of vacancy (Fig. 7g) is rather high and close to the values for graphite. [39]

In the case of non-graphitic defect, the decomposition of activated oxygen (Fig. 7d) with the formation of two epoxy groups (Fig. 7f) is an exothermic process that corresponds to the instability of oxidation of graphene in the vicinity of non-graphitic boron defects. Thus despite a rather low energy cost of the vacancy formation (Fig. 7h) the instability of previous configurations makes the formation of this type of defect rather improbable at ambient conditions. Thus, based on results of DFT modeling we could conclude that formation of defects suitable for the penetration of oxygen atoms to the copper substrate is the non-preferable scenario.

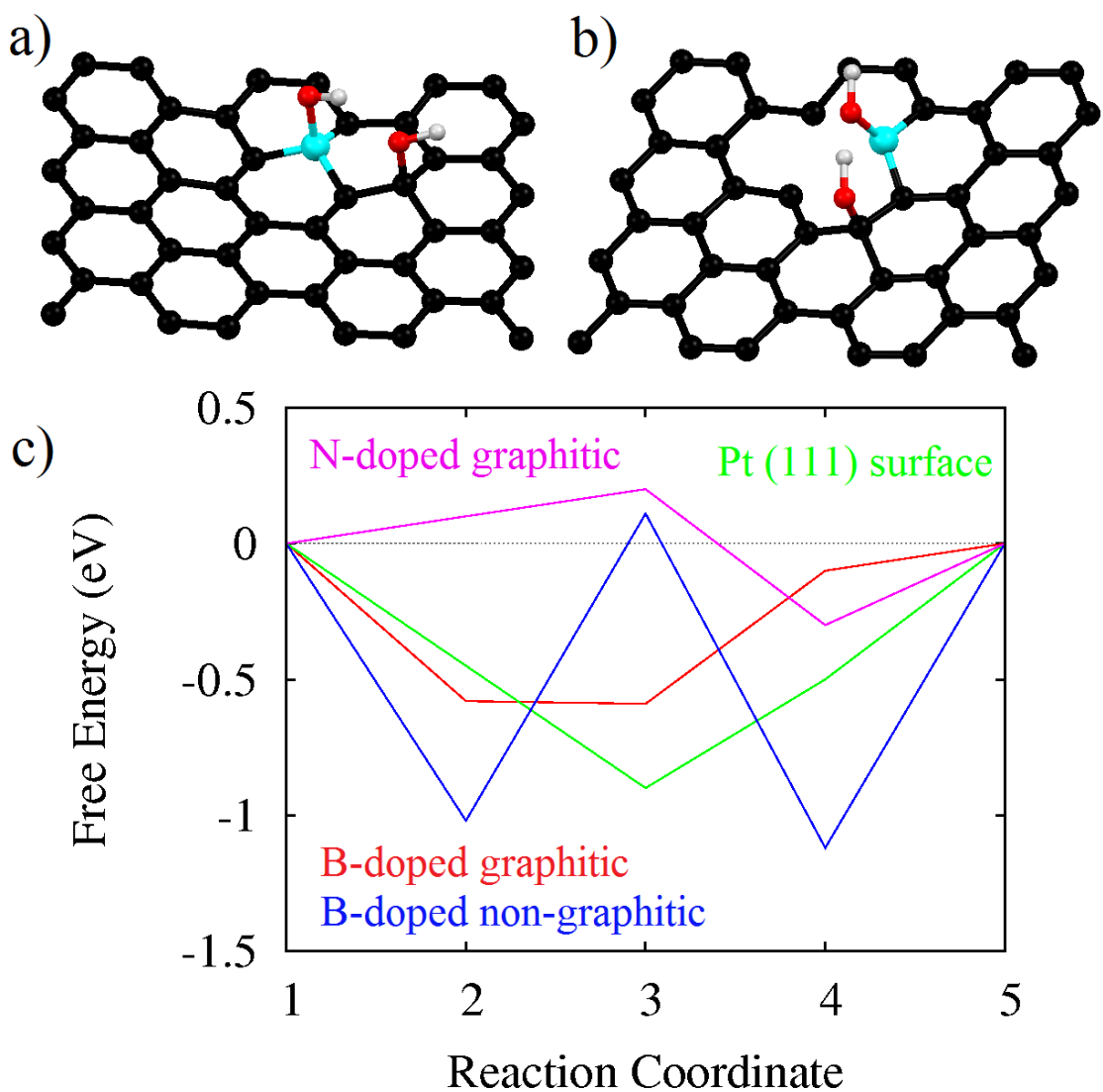

**Figure 8.** Optimized atomic structure of the dormation of two hydroxyl grpups on B-doped graphene on copper with graphitic (a) and non-graphitic defects (b) corresponding to 4rd step of ORR (see description in text), and free energy diagram for ORR over B-doped graphene on copper, N-doped graphene [46] and platinum [47].

## 3.5. Modeling of Catalytic Properties

Described combination of the favorability of molecular oxygen activation with the instability of epoxy groups in the vicinity of non-graphitic boron-defects makes this material prospective candidate for catalysis and requires further evaluation of the catalytic properties. For the evaluation of the feasibility of studied systems we performed the calculations of the free energy of two electrons oxygen reduction reaction (ORR) in acid media. For the graphene this reactions contain five steps [46]:

1) Physisorption of molecular oxygen on the surface (Fig. 7a,b),
2) Activation of the oxygen molecule (Fig. 7c,d),
3) Formation of epoxy groups (Fig 7e,f),
4) Interaction of epoxy groups with hydrogen $2O_{ads} + H_2 \rightarrow 2OH_{ads} + 2e^-$ (Fig. 8a,b),
5) Interaction of hydroxyl groups with hydrogen $2OH_{ads} + H_2 \rightarrow 2H_2O + 2e^-$.

The calculation of the free energy was performed by a previously developed method [47] for platinum based catalysis using the formula: $G = \Delta E_N - neU + E_{ZP}$, where $\Delta E_N$ is the energy difference between total energies at N and N+1 step of reaction processes, e the electron charge, U the equilibrium potential, n is the number of oxygen atoms, and $E_{ZP}$ the zero point energy correction, respectively. The values of U (1.23 V) and zero point energy corrections are the same as the ones used in previous works. [46,47] Results of the calculations (Fig. 8c) demonstrate that B-doped graphene on copper with graphitic defects have catalytic performance a bit worse that nitrogen doped graphene but better than platinum, and B-doped graphene with non-graphitic defects is a bit worse than platinum but have performance better than other transition metals (see discussion in Ref. [47]). The combination of rather low magnitudes of Gibbs free energies of ORR and stability makes B-doped graphene on copper prospective catalytic material. The further progress in the study of the chemical bonding in interfaces and its effect on the stability of electronic configurations of impurity atoms in a graphene coating favorable for corrosion resistance and photocatalytic activity is apparently associated with the calculation of some quantitative parameters in particular interfacial adhesion of doped graphene and metal substrate. It can be realized with help of recently developed bond order-length-strength (BOLS) correlation theory [48-51].

## 4. Conclusions

The measurements of XPS, Raman and OSEE specters evidences that boron atoms substitute carbon atoms which provide the homogeneity of graphene layer and prevents to oxidation of Cu-substrate. DFT calculations confirm this conclusion according to which the probability of creation of carbon vacancy near graphitic defects is rather low because high energy cost and in the case of non-graphitic defect despite favorability of oxygen activation, the next step in the perforation of graphene is unstable. Theoretical modeling also predicts high catalytic performance of B-doped graphene on copper substrate for oxygen reduction reaction.

**Acknowledgements**
We thank Prof. A.M. Rao and his colleagues from Clemson University for preparation of boron-doped graphene/copper samples and Raman spectra measurements. The XPS measurements were supported by Russian Science Foundation (Project 14-22-00004). D.W.B. acknowledges support from the Ministry of Education and Science of the Russian Federation, Project №3.7372.2017/БЧ.

**References**
1. Y. Sun, Q. Wu, G. Shi, Graphene based new energy materials, Energy Environ. Sci. 4 (2011) 1113.
2. L. Dai, Functionalization of graphene for efficient energy conversion and storage, Acc. Chem. Res. 46 (2012) 31.
3. J. Xia, F. Chen, J. Li, N. Tao, Measurement of the quantum capacitance of grapheme, Nat. Nanotech. 4 (2009) 505.
4. M. F. El-Kady, R. B. Kaner, Scalable fabrication of high-power graphene micro-supercapacitors for flexible and on-chip energy storage, Nat. Commun. 4 (2013) 1475.
5. M. F. El-Kady, V. Strong, S. Dubin, R. B. Kaner, Laser scribing of high-performance and flexible graphene-based electrochemical capacitors, Science 335 (2012) 1326.
6. Z. S. Wu, A. Winter, L. Chen, Y. Sun, A. Turchanin, X. Feng, K. Müllen, Three-Dimensional Nitrogen and Boron Co-doped Graphene for High-Performance All-Solid-State Supercapacitors, Adv. Mater. 24 (2012) 5130.
7. Z. Yang, Z. Yao, G. Li, G. Fang, H. Nie, Z. Liu, X. Zhou, X. A. Chen, S. Huang, Investigation of homologous series as precursory hydrocarbons for aligned carbon nanotube formation by the spray pyrolysis method, ACS Nano 6 (2011) 205.
8. R. Lv, M.Terrones, Towards new graphene materials: doped graphene sheets and nanoribbons, Materials Lett. 78 (2012) 209.
9. F. Banhart, J. Kotakoski, A. V. Krasheninnikov, Structural Defects in Graphene, ACS Nano 5 (2011) 26.
10. L. Pan, Y. Que, H. Chen, D. Wang, J. Li, C. Shen, W. Xiao, S. Du, H. Gao, S. T. Pantelides, Room-temperature, low-barrier boron doping of grapheme, Nano Lett. 15 (2015) 6464.
11. S. Iijima, Helical microtubules of graphitic carbon, Nature 354 (1991) 56.
12. K. S. Novoselov, et al. Electric field effect in atomically thin carbon films, Science 306 (2004) 666.

13. Y. Zhang, D. Zhang, C. Liu, Novel chemical sensor for cyanides: boron-doped carbon nanotubes, J. Phys. Chem. B 110 (2006) 4671.
14. J. Dai, J. Yuan, J. P. Giannozzi, Gas adsorption on graphene doped with B, N, Al, and S: a theoretical study, Appl. Phys. Lett. 95 (2009) 232105.
15. B. Biel, X. Blase, F. Triozon, S. Roche, Anomalous doping effects on charge transport in graphene nanoribbons, Phys. Rev. Lett. 102 (2009) 096803.
16. H. Terrones, R. Lv, M. Terrones, M. S. Dresselhaus, The role of defects and doping in 2D graphene sheets and 1D nanoribbons, Rep. Prog. Phys. 75 (2012) 062501.
17. M. Xing, W. Fang, X. Yang, B. Tian, J. Zhang, Highly-dispersed boron-doped graphene nanoribbons with enhanced conductibility and photocatalysis, Chem. Commun. 50 (2014) 6637.
18. Z. -S. Wu, W. Ren, L. Xu, F. Li, H. -M. Cheng, Doped graphene sheets as anode materials with superhigh rate and large capacity for lithium ion batteries, ACS Nano 5 (2011) 5463.
19. C. E. Lowell, Solid solution of boron in graphite. J. Am. Ceram. Soc. 50 (1967) 142.
20. H. Murty, D. Beiderman, E. Heintz, Apparent catalysis of graphitization. 3. Effect of boron, Fuel 56 (1977) 305.
21. L. E. Jones, P. A. Thrower, Influence of boron on carbon fiber microstructure, physical properties, and oxidation behavior, Carbon 29 (1991) 251.
22. R. B. Kaner, J. Kouvetakis, C. E. Warble, M. L. Sattler, N. Bartlett, Boron-carbon-nitrogen materials of graphite-like structure, Mater. Res. Bull. 22 (1987) 399.
23. J. M. Soler, E. Artacho, J. D. Gale, A. Garsia, J. Junquera, P. Orejon, D. Sanchez-Portal, The SIESTA method for ab initio order-N materials simulation, J. Phys.: Condens. Matter 14 (2002) 2745.
24. I.-S. Byun, et al., Electrical control of nanoscale functionalization in graphene by the scanning probe technique. NPG Asia Mater, 6 (2014) e102.
25. D.W. Boukhvalov, Y.W. Son, R.S. Ruoff, Water Splitting over Graphene-Based Catalysts: Ab Initio Calculations, ACS Catal. 4 (2014) 2016.
26. J.P. Perdew, A. Zunger, Self-interaction correction to density-functionalapproximations for many-electron systems, Phys. Rev. B 23 (1981) 5048.
27. M. Vanin, J. J. Mortensen, A. K. Kelkkanen, J. M. Garcia-Lastra, K. S. Thygesen, K. W. Jacobsen, Graphene on metals: A van der Waals density functional study. Phys. Rev. B 81 (2010) 081408(R).
28. O. N. Troullier, J. L. Martins, Efficient pseudopotentials for plane-wave calculations, Phys. Rev. B 43 (1991) 1993.
29. H. J. Monkhorst, J. D. Pack, Special points for Brillouin-zone integrations, Phys. Phys. Rev. B 13 (1976) 5188.
30. L. Bokobza, J. L. Bruneel, M. Couzi, Raman spectra of carbon-based materials (from graphite to carbon black) and of some silicone composites, C J. Carbon Res. 1 (2015) 77.
31. L. Wang, Z. Sofer, P. Šimek, I. Tomandl, M. Pumera, Boron-doped graphene: scalable and tunable p-type carrier concentration doping, J. Phys. Chem. C 117 (2013) 23251.
32. D. Gonzalez-Larrude, Y. Garcia-Basabe, F. L. Freire Jr., M. L. M. Rocco, Electronic structure and ultrafast charge transfer dynamics of phosphorous doped graphene layers on a copper substrate: a combined spectroscopic study, RSC Adv. 5 (2015) 74189.
33. A. G. Shard, Detection limits in XPS for more than 6000 binary systems using Al and Mg Kα X-rays, Surface and Interface Analysis, 46 (2014) 175.
34. O. Kane, Theory of Photoelectric Emission from Semiconductors, Phys. Rev. 127 (1962) 131.
35. V.S. Fomenko. Emission properties of materials. (Ed. by G.V.Samsonov). Naukova Dumka, Kiev (1970) 146.
36. L. Tseteris and S.T. Pantelides, Graphene: An impermeable or selectively permeable membrane for atomic species, Carbon 67 (2014) 58.
37. L. Gao, J. R. Guest, N. P. Guisinger, Epitaxial Graphene on Cu(111). Nano Lett. 10 (2010) 3512.
38. R. He, L. Zhao, N. Petrone, K. S. Kim, M.l Roth, J. Hone, P. Kim, A. Pasupathy, A. Pinczuk, Large Physisorption Strain in Chemical Vapor Deposition of Graphene on Copper Substrates. Nano Lett. 12 (2012) 2408.
39. M. Topsakal, H. Sahin, S. Ciraci, Graphene coatings: An efficient protection from oxidation, Phys. Rev. B 85 (2012) 155445.
40. R. Faccio, L. Fernandez-Werner, H. Pardo, C. Goyenola, O. N. Ventura, A. W. Mombru, Electronic and structural distortions in graphene induced by carbon vacancies and boron doping, J. Phys. Chem. C 114 (2010) 18961.

41. B. Yang, H. Xu, J. Lu, K. P. Loh J. Periodic grain boundaries formed by thermal reconstruction of polycrystalline graphene film, J. Am. Chem. Soc. 136 (2014) 12041.
42. L. R. Radovic, Active Sites in Graphene and the Mechanism of $CO_2$ Formation in Carbon Oxidation, J. Am. Chem. Soc. 131 (2009) 17166.
43. L. Ferrighi, M. Datteo, C. Di Valentin, Boosting Graphene Reactivity with Oxygen by Boron Doping: Density Functional Theory Modeling of the Reaction Path, J. Phys. Chem. C 118 (2014) 223.
44. G. Fazio, L. Ferrighi, C. Di Valentin, Boron-doped graphene as active electrocatalyst for oxygen reduction reaction at a fuel-cell cathode, J. Catal. 318 (2014) 203.
45. S. Agnoli, M. Favaro, Doping graphene with boron: a review of synthesis methods, physicochemical characterization, and emerging applications, J. Mater. Chem. A 4 (2016) 5002.
46. D.W. Boukhvalov, Y.W. Son, Oxygen reduction reactions on pure and nitrogen-doped graphene: a first-principles modeling. Nanoscale 4 (2012) 417.
47. J. Nørskov, J. Rossmeisl, A. Logadottir, L. Lindqvist, J. R. Kitchin, T. Bligaard and H. Jónsson, Origin of the Overpotential for Oxygen Reduction at a Fuel-Cell Cathode. J. Phys. Chem. B 108 (2004) 17886.
48. C.Q. Sun, Y. Sun, Y.G. Nie, Y. Wang, J. S. Pan, G. Ouyang, L.K. Pan, and Z. Sun, Coordination-Resolved C-C Bond Length and the C 1s Binding Energy of Carbon Allotropes and the Effective Atomic Coordination of the Few-Layer Graphene, J. Phys. Chem. C 113 (2009) 16464.
49. Z.S. Ma, Yan Wang, Y.L. Huang, Z.F. Zhou, Y.C. Zhou, Weitao Zheng, Chang Q. Sun, XPS quantification of the hetero-junction interface energy, Appl. Surf. Sci. 265 (2013) 71.
50. W. Gao, P. Xiao, G. Henkelman, K.M. Liechti and R. Huang, Interfacial adhesion between graphene and silicon dioxide by density functional theory with van der Waals corrections, J. Phys. D: Appl. Phys. 47 (2014) 255301.
51. X. Liu, X. Zhang, M. Bo, L. Li, H. Tian, Y. Nie, Y Sun, S. Xu, Y. Wang, W. Zheng, and C. Q Sun, Coordination-Resolved Electron Spectrometrics, Chem. Rev. 115 (2015) 6746.